\documentclass[onecolumn,preprintnumbers,elsart]{revtex4}
\usepackage{eurosym}
\usepackage{amsmath}
\usepackage{booktabs}
\usepackage{cases}
\usepackage{mathrsfs}
\usepackage{graphicx}
\usepackage{dcolumn}
\usepackage{bm}
\usepackage[center]{subfigure}
\usepackage{color}
\usepackage{float}
\usepackage{ulem}
\usepackage{bm}
\usepackage{amsmath}
\usepackage{upgreek}
\usepackage{mathptmx} 
\usepackage{courier}

\begin{document}

\title{Abnormal motion of optical vortex-antivortex-coupled wavepackets in the parabolic potential}
\author{Haolin Lin$^{1,2\dagger}$, Junhui Jia$^{1,2\dagger}$, Chunhao Liang$^{1*}$, Yanwen Hu$^{2}$, Boris A. Malomed$^{3,4}$, Yangjian Cai$^{1*}$ and Shenhe Fu$^{1}$}
\email{ cliang@dal.ca; yangjiancai@sdnu.edu.cn (Y. Cai); fushenhe@sdnu.edu.cn (S. Fu)}

\baselineskip18pt
\address{\baselineskip18pt
$^{1}$Shandong Provincial Key Laboratory of Light Field Manipulation Physics and Applications \& School of Physics and Optoelectronics, Shandong Normal University, Jinan 250014, China \\
$^{2}$Department of Optoelectronic Engineering, Jinan University, Guangzhou
510632, China \\
$^3$Department of Physical Electronics, Tel Aviv University, Tel Aviv 69978, Israel \\
$^4$Instituto de Alta Investigaci\'{o}n, Universidad de Tarapac\'{a}, Casilla 7D, Arica, Chile}

\maketitle 
\noindent
\textbf{\noindent The (quasi)particles or structured wavepackets in parabolic potentials exhibit well-known harmonic oscillations, typically described by the popular Lissajous equations. However, such conventional harmonic laws rely on a fundamental assumption that the different constituent components of the (quasi)particles or wavepackets do not interact. Here we challenge this paradigm, by taking advantage of intrinsic couplings among distinct constituents, specifically, by leveraging nontrivial couplings between vortices and antivortices embedded in a spatially structured wavepacket. We demonstrate theoretically and experimentally abnormal motion by considering two different waveforms. For a vortex-antivortex (VAV)-coupled dipole mode, we reveal counterintuitive propagation regimes, including periodic annihilation and regeneration of the dipole, its non-orbital motion and realization of a critical equilibrium state in the absence of nonlinearity. For a circular chain of vortices with an antivortex set at the center, we successfully tune the oscillation frequency of the overall configuration, thus disobeying the classical Lissajous trajectories, by engineering the nonlocal VAV couplings. The observed VAV braids and the period increase in the harmonic potential can be used in optical information encoding. Owing to the general character of the wave, our results suggest the investigation of anomalous VAV dynamics in other settings, including electronics, acoustics, hydrodynamics and Bose-Einstein condensates.} 

\newpage 
\noindent\textbf{Introduction}
\newline
\noindent Harmonic oscillations of particles or quasiparticles represent a fundamentally important phenomenon across multiple branches of physics, inspiring extensive theoretical and experimental studies \cite{Baker2005} and enabling diverse technological applications \cite{Faller2010,Tosi,Lan2021}. In a harmonic-oscillator potential (HOP), the particle initially displaced from equilibrium position undergoes oscillations at a frequency determined solely by the potential itself \cite{Baker2005,Faller2010,Tosi,Lan2021}. In two dimensions, imparting an initial momentum perpendicular to the potential gradient results in particle motion along the popular Lissajous trajectory \cite{KJ2006}. This particle picture in the HOP extends directly to wave regime, where wavepackets undergo particle-like dynamics, propagating without distortion along elliptic Lissajous orbits \cite{Zhang2015,Moulieras2012}. Based on this wave–particle analogy, a range of wave- and particle-based dynamical behaviors in the HOPs have been uncovered \cite{Li2024, Sonin1987}. These span from atomic and molecular vibrations in quantum systems \cite{Lan2021,Yanez1994, Lorente2000,Gerving2012} to the dynamics of structured wavepackets in classical wave settings \cite{Jia2023}. \\
\indent The long-standing HOP laws rest on a fundamental assumption: the constituent components of the trapped (quasi)particles or wavepackets exhibit no mutual interactions. This premise breaks down, however, in systems where such interactions become significant. In ultracold molecules, for instance, long-range anisotropic dipole–dipole interactions are known to play a role in the formation of quantum phases such as the dipolar Bose–Einstein condensates (BEC) \cite{Niu1999,Boris2008}. In spatially structured wavepackets, the coexistence of helical and anti-helical phase structures generates an extrinsic orbital angular momentum (OAM) \cite{Padgett2002,Zayats2015}, distinct from its intrinsic counterpart \cite{Allen1992,Barnett2002}. These inter-particle or inter-wavepacket couplings alter the dynamical behavior in the HOP, leading to remarkable deviations from conventional Lissajous trajectories. In particular, the vortex–antivortex(VAV)-coupled wavepackets offer a unique mechanism for the transverse control of motion \cite{Bliokh2006,Lin2022,Freilich2010,Liu2025}, enabling the realization of spin–orbit coupling in optical settings \cite{optics-SOC, Zhang2024}. When such VAV wavepackets evolve under the HOP, the interplay between vortics, antivortics, and the spatial confinement is expected to give rise to unconventional dynamical phenomena beyond the classical paradigms. \\
\indent {We emphasize that studies of the optical VAV interactions in the HOP were scarce, both in theory and experiment. There is a plethora of theoretical works that dealt with the propagation dynamics of structured light in the HOP, including the accelerating-vortex beams \cite{Deng2021,Liu2020}, Hermite-Gaussian vortex beams \cite{Chen2019,Wu2020}, Laguerre-Gaussian vortex beam \cite{Hu2019}, cosine-Gaussian beams \cite{Zhang2021}, vector-vortex beam \cite{Jia2023}, partially coherent vortex beam \cite{Jing2020}, and a light beam that contains multiple vortices and antivortices \cite{He2024,Roux2004,Lin2023,Zhan2020aa,Ferrando2023}. However, these works did not address the structured light that carries the interacting vortices and antivortices. As a result, the previously studied propagation dynamics of the structured light follow the conventional laws of motion in the HOP. Experimentally, generating the VAV-coupled wavepacket requires a precise Fourier transform of the complex wave field, which encodes inherent orbit–anti-orbit couplings governing its real-space VAV interactions. Previous techniques \cite{Ferrando2023,Fan2021,Gao2021,Ding2018,Vyas2007}, which neglect these VAV couplings, thus fail to produce the required VAV-coupled states in experiments. As a consequence, the understanding of the VAV-coupled dynamics in the HOP systems remains limited.}  \\
\indent In this article, we report a combined theoretical and experimental investigation of nontrivial VAV couplings and the resulting unconventional dynamics of structured optical wavepackets in the HOP. To uncover these anomalous phenomena, we study two distinct VAV-coupled wavepackets: a fundamental dipole structure and a circular VAV chain encircling a central antivortex. For the dipole structure, the intrinsic VAV couplings give rise to unconventional dynamical regimes, which cannot be predicted by the usual harmonic formulas. These include recursive annihilation and revival of the dipole pair, non-orbital oscillation of the dipole, and linear propagation of a stationary dipole structure, all of which are achieved in the absence of optical nonlinearity \cite{Ye2020,Ye2020a,Boris2011}, in sharp contrast to behaviors unaffected by such VAV couplings. In the circular chain, the presence of the nonlocal VAV couplings allows us to effectively tune the oscillation frequency in the potential, thus disobeying the classical Lissajous paradigm. It may be relevant to mention that, apart from the optical VAV interactions, a similar concept has been addressed in other physical settings, including superconductors \cite{Chaves2011}, BEC \cite{Freilich2010,Croszek2016,Ferlaino2022}, quantum fluids \cite{Tosi,Gauthier2019,Nardin2011}, and the exciton-polariton condensates \cite{Byrnes2014}. However, the VAV dynamics governed by the nonlinear Gross-Pitaevskii equation is quite different from those observed in the present work. Our investigations suggest further studies of other forms of VAV-coupled wavepackets, which may yield intriguing VAV phenomena. Such abnormal dynamical effects open potential applications \cite{Ren2016, Yang2021, Ouyang2021}, particularly in scenarios where traditional harmonic oscillations are absent or insufficient. \\
\textbf{Concept of the VAV couplings}\newline
\noindent We start by reviewing interactions between two tropical cyclones in the atmospheric environment, with opposite winding fields, as schematically illustrated in Fig. 1a. The two reciprocal cyclones are extremely unstable due to their mutual interactions, gradually changing their typhoon eyes (named as the singularity cores) in the course of evolution. These cyclones exhibit the popular Fujiwhara effect \cite{Bjerknes1921,Alpert1995}, which manifests the couplings among them. As these reciprocal cyclones reach a critical distance, their cyclonic circulations begin to overlap: each tropical typhoon undergoes the action of the opposite azimuthal wind field of the other. As a consequence, both cyclones start to translate according to the motion of their bary centers \cite{Bjerknes1921,Alpert1995}. If the tropical cyclones further get close, they can even merge, and then regenerate as new cyclones during their continuous evolutions \cite{Bjerknes1921}. These natural atmospheric phenomena manifest the important cyclone-anticyclone couplings. \\
\indent The optical VAV coupling studied here originates from the intrinsic interplay between phase singularities of opposite topological elements embedded in a paraxially propagating light field. To illustrate this concept, Fig. 1b presents the transverse phase pattern of a VAV-coupled dipole at the propagation origin ($z=0$), visualized through its isophase contours. The dipole is built of two elements with opposite singularity polarities (helicities), separated by spacing $d$, which are symmetrically positioned at transverse points $x=\pm d/2$ and embedded in the Gaussian envelope, $E_{0}\left(x,y\right) =\exp [-(x^{2}+y^{2})/w_{0}^{2}]$, where $w_{0}$ is the Gaussian waist and $\left( x,y\right) $ are the transverse coordinates. The VAV dipole state is therefore modeled by the optical-wave envelope
\begin{equation}
G\left( x,y\right) =E_{0}\left(x,y\right) (u-d/2)(u^{\ast }+d/2)
\end{equation} 
where $u=x+iy\equiv r{\exp (i\phi )}$ and $u^{\ast }=x-iy\equiv r{\exp (-i\phi )}$ carry the helical and antihelical phase structures. \\
\indent The VAV couplings are manifested as follows: first, the isophase contours of the dipole connect the positive and negative phase singularities directly, forming non-closed isophase curves, see Fig. 1b. Second, the spatially-patterned arrows (Fig. 1e), which reveal the phase gradient of the dipole, suggest that the dipole exhibits upward motion trend in the phase-coupling field, as indicated by the arrows marked by a red rectangle. These observations imply the presence of nontrivial VAV couplings. As the propagation proceeds, the dipole moves upward and the two elements approach each other, as indicated by the isophase contours (Fig. 1c) and the corresponding phase gradient (Fig. 1f). The arrows marked by the red rectangle suggest that the dipole keeps moving upward until the two elements collide and annihilate, resembing the cyclone–anticyclone couplings in geophysical flows \cite{Bjerknes1921, Alpert1995}. Following the annihilation, the previously non-close isophase curves become closed, signifying disappearance of the topological phase structure (Fig. 1d). Consequently, the wavefront loses its helical character, as indicated by the tanglesome arrows (Fig. 1g).  
\begin{figure}[t]
	\centering
	\includegraphics[width=17cm]{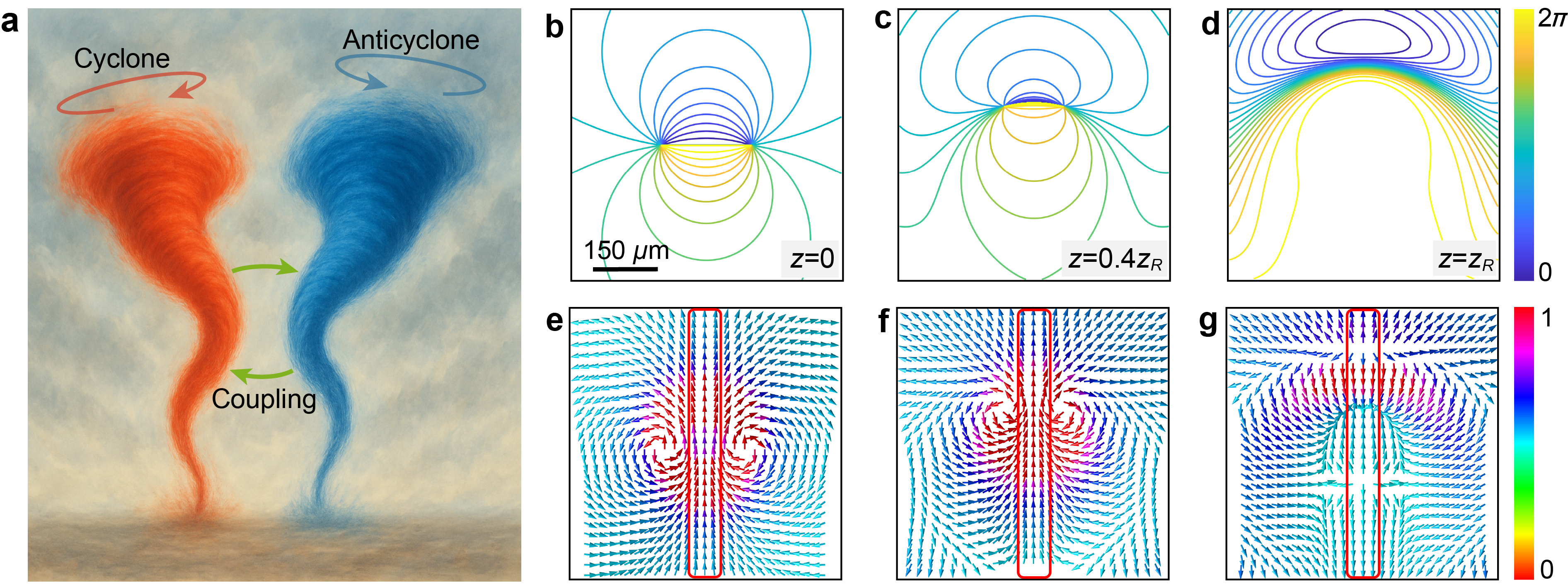}
	\caption{\baselineskip14pt \textbf{Coupling phenomena between the vortex and antivortex}. \textbf{a}, An analogous coupling between the cyclone and the anticyclone in the atmospheric environment. These tropical cyclones can merge and revive, manifesting the cyclone-anticyclone couplings. \textbf{b}-\textbf{d}, Coupling effects in the dynamical VAV dipole, manifested by its isophase contours recorded at: \textbf{b}, $z=0$; \textbf{c}, $z=0.4z_R$; and \textbf{d}, $z=z_R$. \textbf{e}-\textbf{g}, The corresponding phase gradients to \textbf{b}-\textbf{d}. The colorbar in panels \textbf{e}-\textbf{g} denotes the strength of the phase gradient. Panels \textbf{b}-\textbf{g} share identical scale bar: 150 $\mu$m. }
\end{figure}\newline
\indent The dynamical VAV couplings can be understood by tracking how their phase singularity positions change as they propagate. This motion trajectory can be derived by means of the framework elaborated in \cite{PRA2021}, written as: 
\begin{equation}
x_{\pm}(z)=\pm\left(d+\Delta x\right),\quad y_{\pm}(z)=dz/z_R+\Delta y
\end{equation}
where ($x_{+}$, $y_{+}$) and ($x_{-}$, $y_{-}$) are the transverse positions of the vortex and antivortex, respectively. The terms $\Delta x=[1/(2\pi^2dw_0^2)-1/(8\pi^2d^3)]\lambda^2z^2$ and $\Delta y=-\lambda z/(2\pi d)$ represent transverse shifts in their positions, caused by the VAV couplings. In Eq. (2), $z_R=\pi w_0^2/\lambda$ denotes the Rayleigh diffraction range, with $\lambda$ being the carrier wavelength. This VAV coupling is propagation-dependent, resulting in $z$-dependent $\Delta x$ and $\Delta y$. When $\Delta x=0$ and $\Delta y=0$, there is no coupling between the elements. This happens when the structure contains only vortices (or only antivortices) with identical topological charges \cite{Haolin2019}. \\
\textbf{Abnormal dynamics of the VAV-coupled dipole}\newline
\noindent We then reveal abnormal behaviors of the VAV-coupled dipole in the HOP, as schematically shown in Fig. 2a, which gives rise to a phase diagram of the dipole motion, as shown in Fig. 2b. We explain this intriguing phase diagram below. Theoretically, the potential is modeled as $V(x,y)=-(1/2)n_{0}k_{0}\alpha ^{2}(x^{2}+y^{2})$, where $n_{0}$ is the ambient refractive index, $k_0=2\pi/\lambda$ is the vacuum wave number and $\alpha ^{2}$ determines the potential strength. This leads to a HOP pitch, $p = 2\pi/\alpha$ \cite{Jia2023}. Propagation dynamics of the embedded VAV dipole is governed by the linear Schr\"{o}dinger equation \cite{Segev2007,Zong2016},
\begin{equation}
	\left[ i\frac{{\partial }}{{\partial z}}+\frac{1}{{2{k_{0}}{n_{0}}}}{\nabla
		_{\text{T} }^{2}}+V\left( {x,y}\right) \right] G=0  \label{eq:PWE}
\end{equation}
where ${\nabla _{\text{T} }^{2}}=\partial _{x}^{2}+\partial _{y}^{2}$ represents paraxial diffraction of light. With the initial dipole mode, a solution of Eq. (\ref{eq:PWE}) is obtained as
\begin{equation}
	G(x,y,z)=E(x,y,z)\left[ (Tu+d/2)(Tu^{\ast }-d/2)+B\right]  \label{G}
\end{equation}
where $E(x,y,z)=-in_{0}\alpha B(\lambda \sin \tau )^{-1}\exp [-w(z)({{x^{2}}+{y^{2}}})]$ is the propagation-dependent Gaussian envelope with $\tau \equiv \alpha z$, and the respective complex variable waist is $w(z)=w_{0}^{-2}\left( {\cos \tau}+i\sigma ^{-1}\sin \tau \right) /\left( {\cos \tau +i\sigma \sin \tau }\right) $, with $\sigma ^{-1}\equiv (1/2)n_{0}k_{0}\alpha w_{0}^{2}$. It means that the light envelope $E(x,y,z)$ carrying the VAV dipole maintains its Gaussian profile during the evolution in the HOP. The Gaussian waist becomes a periodic function of $z$ under the action of the potential. {Complex coefficients $B$ and $T$ are expressed as: $B=2i\sin\tau(n_0k_0\alpha)^{-1}/({\cos \tau +i\sigma \sin \tau })$ and $T=({\cos \tau +i\sigma \sin \tau})^{-1}$, respectively. These two parameters $B$ and $T$ become fundamentally important, as they uniquely describe how the VAV dipole evolves.} \\
\indent First, the parameter $T$, which appears in term $\left( {Tu+{d/2}}\right) \left( {T{u^{\ast }}-d/2}\right)$ in Eq. (4), reflects the uncoupled propagation of the dipole elements. This term is a product of two multipliers, each linked to the evolution of the vortex and antivortex components of the dipole. Their respective position vectors $\mathbf{r}_{\pm }(z)$ evolve according to the relation: $\mathbf{r}_{\pm }(z)\equiv \lbrack x_{\pm}(z);y_{\pm}(z)]=\mathbf{V}_{\pm}\cdot \mathbf{r}_{\pm }\left( {0}\right)$. Here ${\mathbf{V}_ \pm }(\tau) = \left[ {\cos \tau, \mp \sigma \sin \tau; \pm \sigma \sin \tau,\cos \tau} \right]$ is a rotational transformation matrix derived from the parameter $T$ (detailed derivation refers to the methods). This matrix captures the conventional orbital motion of the vortex and antivortex under the influence of the HOP, resembling the conventional behavior observed in scenarios without the VAV couplings. Figure S1 further illustrates such normal orbital motion of the elements in the HOP, see Sec. A in the supplementary material. The ellipticity of the orbital trajectories is governed by the parameter $\sigma$, as embedded in the rotational matrix ${\mathbf{V}}_{\pm}$. Consequently, the $T$-related term does not describe any coupling between the vortex and antivortex elements; rather, it underscores their independent evolution in the HOP, producing the Lissajous figures.  \\
\indent Second, the parameter $B$ in Eq. (4) embodies the nontrivial VAV couplings, imposed by the harmonic potential. To clarify its role, we expand $B$ in a Taylor series with respect to the modulation strength $\alpha$. The expansion starts from the zero-potential term
{
\begin{equation}
{B_{0}}=\frac{2i{w_{0}^{2}} z}{k_{0}n_{0}{w_{0}^{2}}+2iz}
\end{equation}}
which describes the intrinsic VAV coupling that modifies transverse motions of the dipole elements during their evolutions. The remainder, $B_1=B-B_0$, captures higher-order corrections in $\alpha$ and represents the coupling between the dipole and the external potential. Detailed derivation for the $B_0$ and $B_1$ refers to the methods. Thus, $B$ simultaneously encodes both the internal VAV interaction and the external dipole–potential coupling, playing a fundamental role in modifying the transformation matrix ${\mathbf{V}_ \pm }(\tau)$. Obviously, it produces a non-negligible modification of the evolution trajectory, which cannot be produced by the standard laws of motion in the HOP. The resulting anomalous trajectories of the dipole’s elements can be derived as
\begin{equation}
\begin{array}{l}
	{x_{\pm }}\left( z\right) =\mp {w_{0}}{\left[ {{\Delta ^{2}}+(1-\Delta^{-2}/4-\Delta^2\sigma^{-2}){{\sigma ^{2}}}{{\sin }^{2}}\tau }\right] ^{\frac{1}{2}}} \\
	{y_{\pm }}\left( z\right) =0.5{w_{0}/\sigma\left( {2\Delta -{\Delta ^{-1}}}\right)\sin \tau }
\end{array}
\end{equation}
where $\Delta =d/(2w_{0})$. Because the parameter $B$ depends on $\tau$, the VAV couplings in the potential is propagation-varying, making the motion of the dipole's elements sensitive to their initial spacing. 
\begin{figure}[t]
	\centering
	\includegraphics[width=13cm]{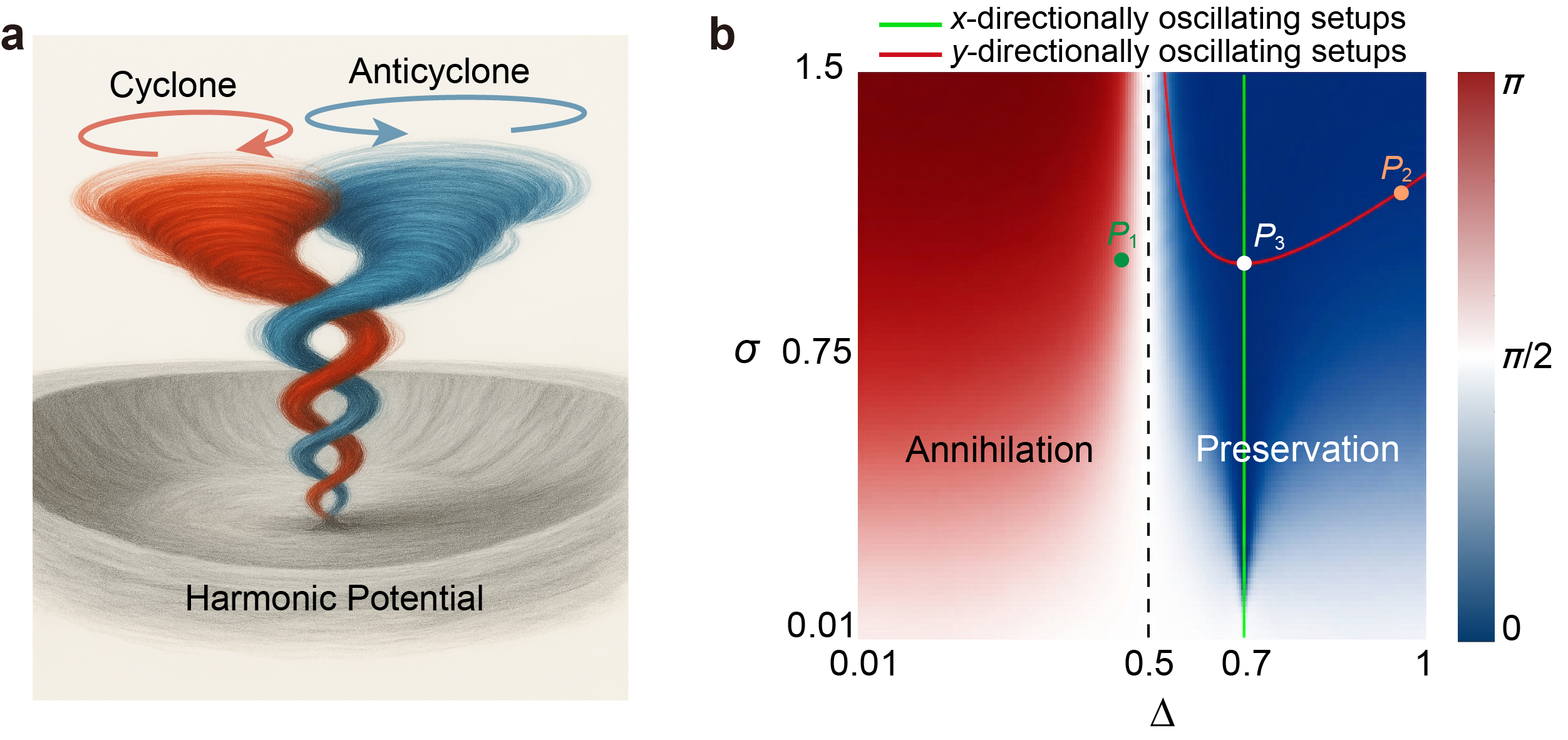}
	\caption{\baselineskip14pt \textbf{The phase diagram of the VAV dipole motion}. \textbf{a}, A similar cyclone-anticyclone-coupled structured trapped in the parabolic potential, emulating the abnormal oscillation. \textbf{b}, An optical VAV dipole trapped in the HOP, leading to different states of motion. The black dashed line, red and green curves depict three different transition boundaries of the VAV dipole state. Points $P_1$, $P_2$, and $P_3$ denote three typical states, which are confirmed by the following experiments. }
\end{figure}
\begin{figure}[t]
	\centering
	\includegraphics[width=15cm]{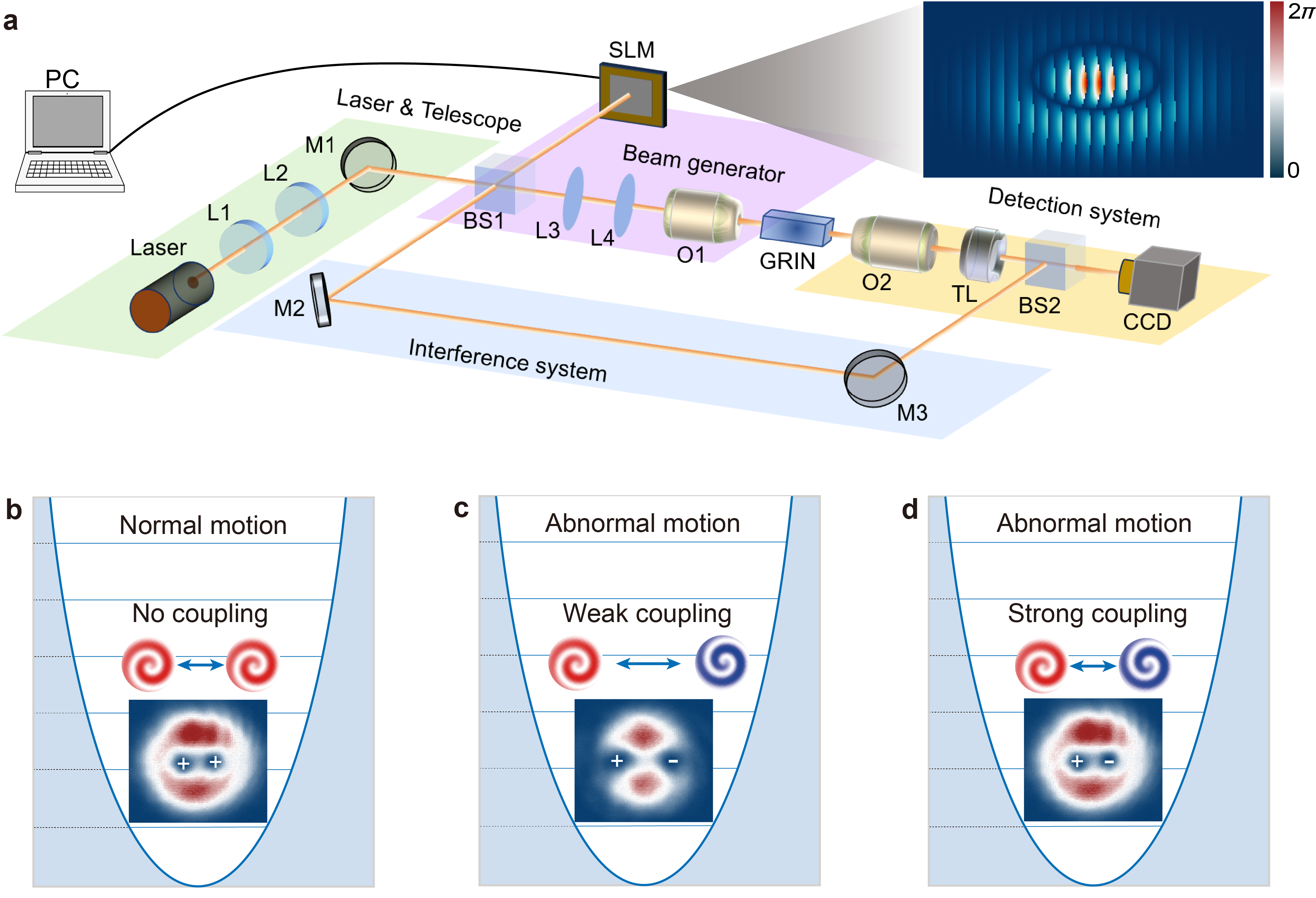}
	\caption{\baselineskip14pt \textbf{VAV dipole in the HOP}.  \textbf{a}, An experimental setup for generating the VAV wavepacket that exhibits nontrivial VAV couplings. L: lens; BS: beam splitter; M: mirror; O: objective lens; TL: tube lens; CCD: charge-coupled device; SLM: spatial light modulator; PC: personal computer. The inset represents a phase-only hologram for the VAV dipole. \textbf{b}, A vortex-vortex wavepacket trapped in the HOP exhibits no intrinsic coupling effect. \textbf{c}, \textbf{d}, A VAV dipole trapped by the HOP manifests weak (\textbf{c}) and strong (\textbf{d}) couplings, which depend on the spacing between the two elements. The insets in panels \textbf{b}-\textbf{d} depict the experimentally generated light fields, featuring the expected structures. The symbols '+' and '-' represent polarities (helicities) of the phase singularities. The red and blue spiral symbols represent vortex and antivortex. }
\end{figure}
\begin{figure}[t]
\centering
\includegraphics[width=18cm]{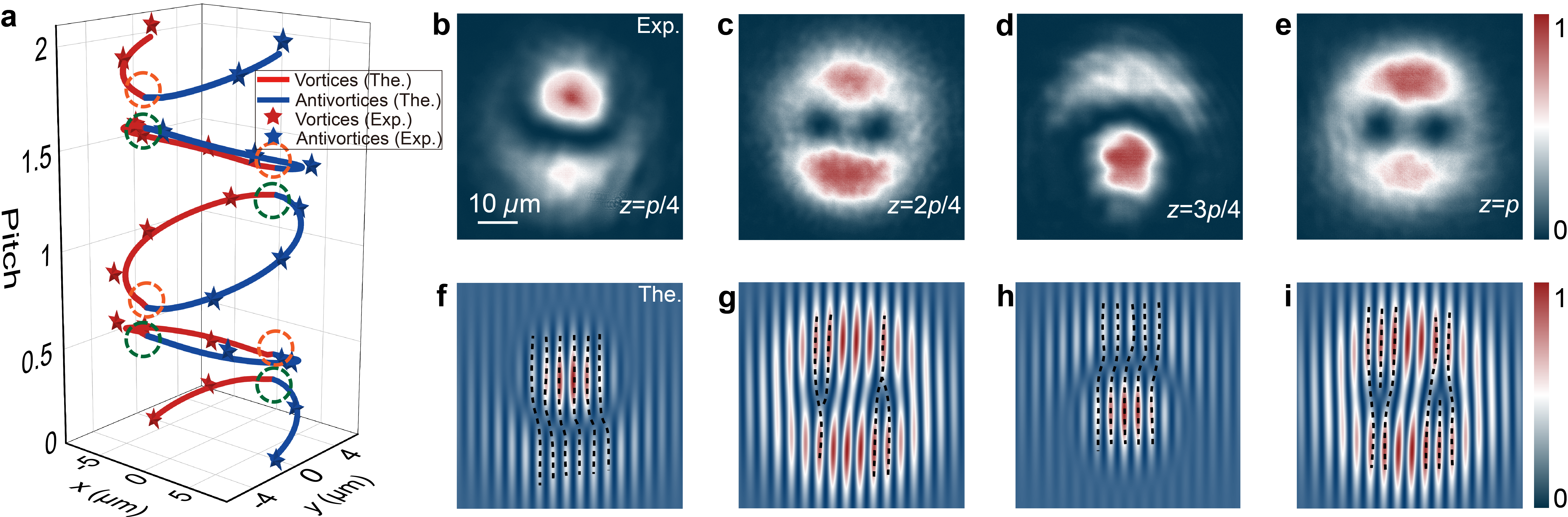}
\caption{\baselineskip14pt \textbf{Observation of recursive annihilation and revival of the VAV dipole}. \textbf{a}, Evolution trajectories of the dipole elements in the HOP. The curves denote the theoretical result, whereas the discrete stars represent the experimental outcome. The green and orange circles highlight the positions where dipole annihilates and revives. \textbf{b}-\textbf{e}, The experimentally recorded light fields at four different propagation distances: \textbf{b}, $z=p/4$; \textbf{c}, $z=p/2$; \textbf{d}, $z=3p/4$; and \textbf{e}, $z=p$. \textbf{f}-\textbf{i}, The simulated interference patterns, corresponding to panels \textbf{b}-\textbf{e}. In this case, the spacing between the dipole elements is set as $\Delta =0.48$, suggesting strong vortex-antivortex coupling. Panels \textbf{b}-\textbf{i} share identical scale bar: 10 $\mu$m. }
\end{figure}\newline
\indent Next, we analyze the expression $x_{\pm}(z)$, which features the distance between the two dipole elements. Due to the periodicity of the system, we consider a propagation distance of one pitch (period) of the potential. The dipole elements are able to reach a minimum distance when the distance approaches to the condition $\sin^2\tau=1$. In this critical scenario, we define a parameter $C_x=\sigma^2(1-\Delta^{-2}/4)$, which determines whether the dipole elements annihilate (when $C_x>0$) or persist ($C_x<0$) during the evolution. If the elements persist, their transverse positions vary as functions of $\sigma$, $\Delta$ and $\tau$. \\
\indent We then address the expression $y_{\pm}(z)$, which shows how the dipole elements vary (oscillate) in the HOP. If $2\Delta-\Delta^{-1}=0$, the dipole state attains an equilibrium state, keeping its position constant; while $2\Delta-\Delta^{-1}\neq0$ indicates an oscillating state. Within these considerations, we define another parameter, $C_y=(2\Delta-\Delta^{-1})^2$, which fully characterizes these states. \\
\indent To figure out a phase diagram that can represent all the states of motion of the VAV dipole, we introduce a new quantity, $C_t(\Delta, \sigma)=\arg(C_x+iC_y)$, where arg($\cdot$) denotes an argument of the complex value. With this definition, we plot the value of $C_t$, with the corresponding phase diagram displayed in Fig. 2b. The phase diagram exhibits three different transition boundaries, marked by the dashed black, red and green lines. The first boundary (dashed line), located at $C_t=\pi/2$, separates the phase diagram in two main regimes, in which the dipole elements annihilate or persist (the left and right regions, respectively). Then, the red curve showcases a boundary of the region in which the two dipole elements keep their $x$ coordinate constant, while oscillating along the $y$ direction. The vertical green curve represents a critical state, which makes the coordinate $y_{\pm}(z)$ of the dipole elements constant, causing the dipole to oscillate along the $x$ coordinate. The intersection point ($P_3$) of the two curves indicates an equilibrium state of the system. \\
\indent We carried out experiments to observe the different phase regimes. This requires to derive the Fourier transform of the dipole, which features the VAV coupling and is used for phase-only encoding (see Methods for detailed derivation and encoding). Figure 3a illustrates an experimental setup, which includes the VAV beam generator, detection system and interference system. A linearly polarized He-Ne laser with an operating wavelength of $\lambda$= 632.8 nm is carefully expanded and collimated using a combination of two optical lenses (L1 and L2). We use a beam splitter (BS1) to divide the incident laser beam into two: a reference laser beam for wavefront characterization and a signal beam, modulated by a phase-only spatial light modulator (SLM) (Holoeye LETO II SLM, 1920$\times$1080). The hologram is produced by means of a technique proposed by Bolduc \cite{Bolduc2013}, as detailed in the Methods section. A phase-only hologram that encodes the VAV dipole configuration is presented in the insert of Fig. 3a. The spatially patterned signal beam, reflected from the SLM, propagates through a 4f system, which is built of two identical lenses (L3 and L4). At the focal plane of L3, an iris is utilized to select the first-order diffraction component of the hologram, while other diffraction components are blocked. The 4f system allows us to transfer the SLM-mediated pattern to the front focal plane of an objective lens (O1, numerical aperture NA=0.17), which performs the Fourier transform of the hologram and generates the target light field at the back focal plane of O1. The generated dipole structure is then normally injected into a GRIN crystal, which emulates the HOP. Parameters of the potential are provided as $\alpha =6.06$ cm$^{-1}$ and $n_{0}=1.611$ \cite{Feit1978}. In the detection part, we consider another objective lens (O2, NA=0.23), together with a tube lens (TL) and a charge-coupled device (CCD), to measure the output wavepacket emerging from the potential. To characterize the wavefront, the signal beam is superimposed with the reference laser beam, generating the plane-wave interference fringes. Figure 3b illustrates the generated elements with identical topological charge, featuring no coupling among them; while Fig. 3c and 3d plot the dipole elements, featuring weak (large $\Delta$) and strong (small $\Delta$) couplings among them, respectively, which result in abnormal motion. The polarity ('+' or '-') of these elements can be identified by the interference patterns (see Fig. S2 in Sec. B of the Supplementary). \\
\indent We start our investigations with a small $\Delta =0.48$ ($d=14\ \mathrm{\mu }$m), where the intrinsic VAV coupling is relatively strong. The phase state is located at point $P_1$ of the phase diagram (Fig. 2b). In this phase regime, the vortex and antivortex exhibit prominent interaction, which gives rise to the result displayed in Fig. 4a. First, the vortex and antivortex attract each other and undergo annihilation in the course of their earlier evolution, due to the attraction between them, resembling the conventional VAV pair evolution in free space \cite{PRA2021}. Second, the dipole surprisingly revives, which is quickly followed by the secondary annihilation, in sharp contrast to the conventional wisdom stating that the VAV dipole cannot be reborn once its elements collide and annihilate \cite{PRA2021}. Indeed, the formula in Eq. (2) cannot reveal the revival of the VAV dipole in the course of its evolution. In our case, the dipole regenerates once again and gradually returns to its original position under the action of the HOP. This anomalous dipole behavior is verified by measuring trajectories of the dipole pivots as depicted in Fig. 4a, where locations of the dipole's annihilation and rebirth are highlighted by the green and orange dotted circles, respectively. Figure 4b-4e depict the experimentally measured intensity distributions of the light fields at different propagation distances, further confirming attraction, annihilation and revival of the dipole elements. Figure 4f-4i illustrate their corresponding interference patterns, showing the locations of the phase singularities. These observations show a good agreement with the theoretical results (see Fig. S3 in Sec. C of the supplementary). \\
\indent Next, the VAV dipole is manipulated to oscillate along a particular direction, manifested as synchronous oscillations of the dipole elements, see a typical point $P_2$ of the phase diagram. This phase regime is different from the conventional orbital motion described by the Lissajous trajectories \cite{KJ2006,Jia2023}. To produce it, we increase the initial spacing to $\Delta =0.93$ ($d=26\ \mathrm{\mu }$m), which relatively weakens the VAV coupling and realizes the critical condition: $\Delta ^{2}\sigma ^{-2}+\Delta ^{-2}=4$. As a result, the $x$ coordinate of the vortex and antivortex pivots remains fixed at ${x_{\pm }} \left( z\right) =\mp {d/2}$, preventing annihilation among the reciprocal elements, whereas the $y$ coordinate follows sinusoidal trajectories without a phase mismatch. Figure 5a showcases this anomalous vertical oscillation both experimentally (data points) and theoretically (curves). The oscillation period exactly matches to the pitch $p$ of the potential. Figure 5b-5e illustrate the measured intensity patterns of light field at four characteristic distances, revealing that the dipole preserves its structure while oscillates along the vertical direction. The polarity (plus or minus) of the dipole elements is indicated in the panels, which can be identified from the interference patterns, see Fig. S4a-S4d in section D of the supplementary. In addition to the vertical oscillation, we can realize a dipole oscillation along horizontal direction, namely, the dipole elements simultaneously oscillate along with the $x$ axis, by setting a critical condition of $\Delta =1/\sqrt{2}$ in Eq. (6). This corresponds to the phase regime, marked by the green line in Fig. 2b.  
\begin{figure}[t]
	\centering
	\includegraphics[width=18cm]{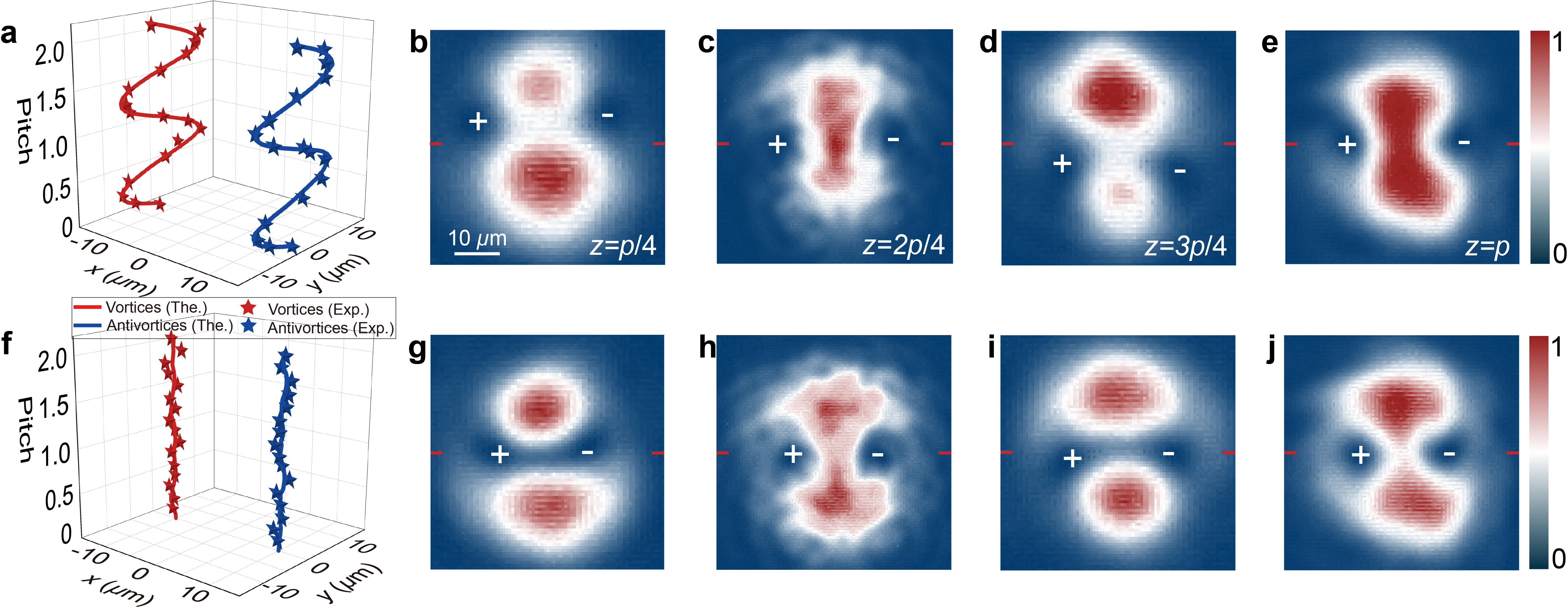}
	\caption{\baselineskip14pt \textbf{Manipulating dipole dynamics using the unique VAV couplings}. \textbf{a}, The VAV dipole is carefully manipulated to oscillate along $y$ axis, by setting $\Delta=0.93$. The curves and data points illustrate theoretical and experimental evolution trajectories of the dipole elements. Red and blue for the vortex and antivortex elements. \textbf{b}-\textbf{e}, The measured light fields at four propagation distances for \textbf{a}: \textbf{b}, $z=p/4$; \textbf{c}, $z=p/2$; \textbf{d}, $z=3p/4$; and \textbf{e}, $z=p$. \textbf{f}, The VAV dipole reaches a critical equilibrium state, by setting $\Delta=\sqrt{2}/2$. As a result, the evolution trajectories become straight lines. Curves and data points illustrate theoretical and experimental results, respectively. \textbf{g}-\textbf{j}, The measured light fields at four propagation distances for \textbf{f}: \textbf{g}, $z=p/4$; \textbf{h}, $z=p/2$; \textbf{i}, $z=3p/4$; and \textbf{j}, $z=p$. Panels \textbf{b}-\textbf{e} and \textbf{g}-\textbf{j} share identical scale bar: 10 $\mu$m. }
\end{figure}\newline
\indent We are able to realize an equilibrium regime (see point $P_3$ in Fig. 2b), in which the individual vortex and antivortex elements are kept by the VAV coupling in the combination with the potential. This showcases sharp contrast to a general property that an antivortex and a vortex simultaneously nested in a wave field tend to annihilate each other, leading to destroying their original arrangements. In this particular regime, transverse motion of the elements disappears, making their evolution trajectory along a straight line, as depicted in Fig. 5f. Such an equilibrium dipole state requires to simultaneously satisfy two critical conditions: $\Delta^{2}\sigma ^{-2}+\Delta ^{-2}=4$ and $2\Delta -\Delta ^{-1}=0$, as suggested by the Eq. (6). This leads to a specific value of the Gaussian waist, $w_{0}=14.2\ \mathrm{\mu }$m, and a critical spacing between the pivots, $\Delta =\sqrt{2}/2$ (tantamount to $d=20\ \mathrm{\mu }$m). With these critical settings, Figure 5f illustrates both experimentally and theoretically the transverse positions of these two elements in the course of their propagations up to two pitches, suggesting a generation of this equilibrium state. Note that the equilibrium state does not require an inclusion of the optical nonlinearities \cite{Ye2020,Ye2020a,Boris2011}. To further confirm this equilibrium state, Fig. 5g-5j present the experimentally measured intensity distributions of the light field at distinct propagation distances, showcasing the nearly propagation-invariant VAV dipole state. The pivots, whose polarities can be identified from the measured interference patterns (see Fig. S4e-S4h in the supplementary), maintain their original positions unchanged during propagation. This unique equilibrium state arises from a precisely balanced vortex-antivortex-potential coupling. \\
\noindent \textbf{Abnormal oscillations with tunable oscillation frequencies}\\
\noindent A general perspective holds that the oscillation frequency of a (quasi)particle or wavepacket in the HOP is solely determined by the potential itself. Once the HOP is established, this oscillation frequency is considered fixed and unalterable. Here we demonstrate that the oscillation frequency of the VAV wavepacket can, in fact, be effectively tuned, thus disobeying the usual harmonic laws. By analyzing a VAV wavepacket with its waveform possessing $N$-fold rotational symmetry, we uncover this unusual dynamics, showing ability to tune the oscillation period. 
\begin{figure}[b]
\centering
\includegraphics[width=17.5cm]{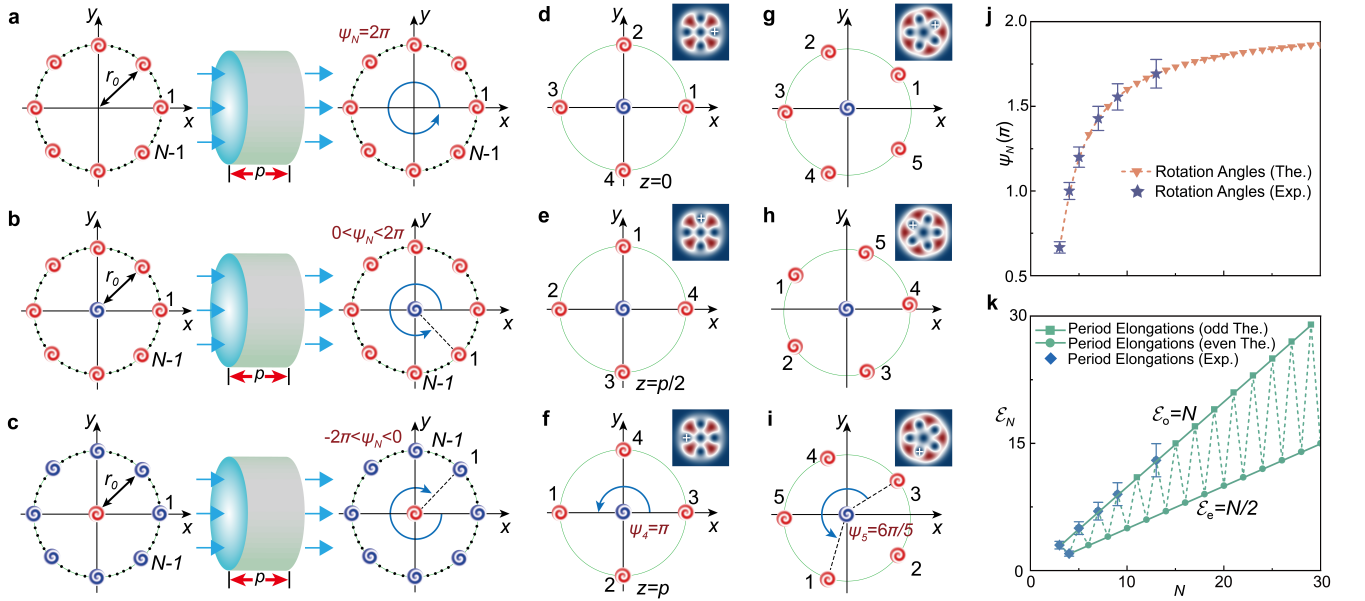}
\caption{\baselineskip14pt \textbf{Abnormal oscillations of the VAV chain with tunable oscillation frequency}. \textbf{a}, Normal oscillation in a case where the structure does not feature VAV couplings. \textbf{b} and \textbf{c}, Abnormal oscillations in cases where the structures feature nonlocal VAV couplings. In panels \textbf{a}-\textbf{c}, red and blue spiral symbols denote the vortex and antivortex. \textbf{d}-\textbf{f}, Transverse distributions of the elements in the four-fold rotationally symmetric configuration at different distances: \textbf{d}, $z=0$; \textbf{e}, $z=p/2$; and \textbf{f}, $z=p$. \textbf{g}-\textbf{i}, The same as described in panels \textbf{d}-\textbf{f}, but in the case of five-fold rotationally symmetric configuration. \textbf{j}, The relationship between the rotation angle $\psi_{N}$ and the order of rotational symmetry $N$ after one-pitch propagation: orange dashed curve and blue discrete stars represent the theoretical and experimental results, respectively. \textbf{k}, The relationship between the increase factor $\mathcal{E}_{N}$ and $N$. } 
\end{figure}\newline
\indent Specifically, the considered VAV wavepacket includes an antivortex at the center ($x=0,y=0$) and $N$ identical vortices symmetrically distributed along a circle of radius $r_{0}$ around the antivortex, with pivots placed at points ${r_{n}}=x_{+}^{(n)}+iy_{+}^{(n)}={r_{0}}\exp [i(2\pi n/N+\theta _{0})]$ ($n=1,2,\ldots ,N$), where $\theta _{0}$ is an angular offset, used to adjust the coordinates of the first vortex, see Fig. 6 for illustrations. The dynamics of this complex VAV wavepacket is governed by the Schr\"{o}dinger equation [Eq. (3)]. We analytically derive a solution of Eq. (3) by using this complex ansatz, written as
\begin{equation}
G(x,y,z)=E(x,y,z)\left\{ ({T{u^{\ast }}})[{{\left( {Tu}\right) }^{N}}-r_{0}^{N}\exp(iN\theta _{0})]+B(Tu)^{N-1}\right\} \label{eq:RotSym}
\end{equation}%
The solution includes the uncoupled term $({T{u^{\ast }}})[{{\left( {Tu}\right) }^{N}}-r_{0}^{N}\exp (iN\theta _{0})]$, which accounts for the independent transverse motion of the outer vortices in the presence of the potential, and a coupling term $B(Tu)^{N-1}$, which nonlocally couples different elements and considerably affects their motions in the system. If the central element is removed (see Fig. 6a), the action of $B(Tu)^{N-1}$ on the vortex elements disappears. Thus, the set of the outer elements rotates as a whole during evolution, returning to the original positions after one-pitch propagation. By contrast, Fig. 6b illustrates that the central antivortex contributes to the nonlocal VAV couplings, described by the term $B(Tu)^{N-1}$ in Eq. (\ref{eq:RotSym}), which affects the orbital motion of the structure. The outer vortices rotate slower in the presence of the VAV couplings, causing an increase of the rotation period, described as $\mathcal{E}_{N}\cdot p$, with the period-increase factor $\mathcal{E}_{N}$ specified below. Reversing the sign of the vortices and antivortices in the structure leads to an opposite rotation with identical velocity magnitude, as illustrated in Fig. 6c. \\
\indent We investigate the period-increase factor $\mathcal{E}_{N}$, by examining motion of the outer vortices under the action of term $B(Tu)^{N-1}$, over a single-pitch propagation. This leads to a corresponding rotation angle $\psi _{N}$ of the overall VAV chain. To understand the relation between $\mathcal{E}_{N}$ and $\psi_{N}$, we assign a position index to each element, as depicted in Fig. 6a-6c. Let us take the Fig. 6b as an illustration. Due to the VAV couplings, the $N$-fold rotationally symmetric configuration rotates in the HOP, which is tantamount to replacing the element at position $r_{N-1}$ by the one at the original position $r_{1}$. The permutation between the two elements gives rise to a relationship between angle $\psi_{N}$ and $N$: ${\psi _{N}}=2\pi (N-2)/N$. This rotation angle, closely associated with the rotational symmetry order $N$, is less than $2\pi $. Figure 6j shows that, as $N$ increases, $\psi _{N}$ grows incrementally, eventually approaching $2\pi $ when $N$ is sufficiently large (cf. Fig. 6a in the absence of the VAV coupling). Indeed, the effect exerted by the central antivortex subsides as the number $N$ of the vortices in the outer circle grows. The resulting VAV structure is therefore equivalent to the initial one rotated by an angle of $\psi_{N}$, namely, $\textbf{r}(z=p)=\textbf{R}(\psi_{N})\cdot\textbf{r}(z=0)$, where $\textbf{R}(\psi_{N})=[\cos(\psi_{N}), -\sin(\psi_{N}); \sin(\psi_{N}), \cos(\psi_{N})]$ denotes the rotational transformation matrix. It satisfies the condition $[\mathbf{R}\left( {{\psi _{N}}}\right) ]^{\mathcal{E}_{N}}=\mathbf{I}$, where $\mathbf{I}$ is the 2$\times$2 identical matrix. This means that, after propagation over distance ${\mathcal{E}_{N}}\cdot p$, the evolving VAV wavepacket returns to its original waveform. Using this relationship, we obtain the period-increase factor as ${\mathcal{E}_{N}}={{\text{LCM}}\left( {N-2,N}\right)}/\left(N-2\right)$, where LCM$(N-2,N)$ is the least common multiple between $N-2$ and $N$.
\begin{figure}[b]
\centering
\includegraphics[width=16cm]{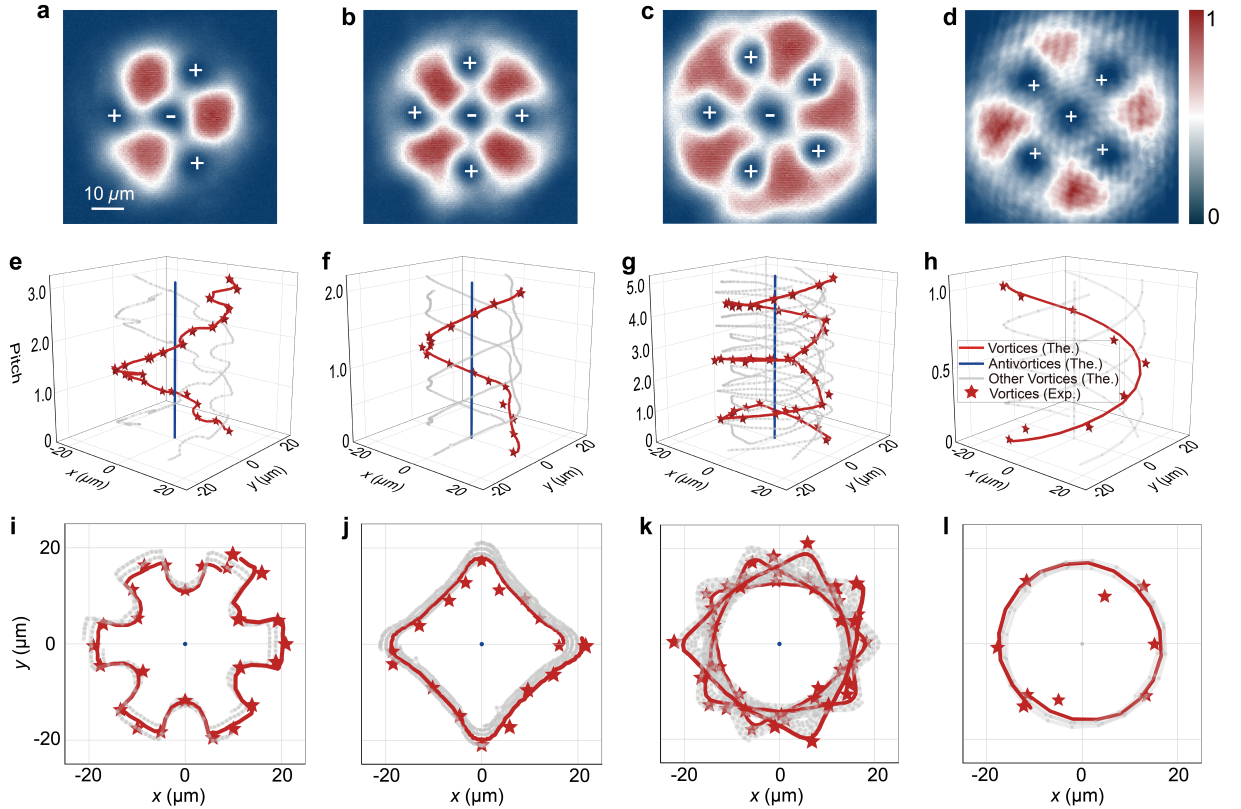}
\caption{\baselineskip14pt \textbf{Observation of period increase in the HOP}. \textbf{a}-{\textbf{d}}, The measured light fields at $z=0$, featuring different VAV configurations characterized by order $N$: \textbf{a}, $N=3$; \textbf{b}, $N=4$; \textbf{c}, $N=5$; and \textbf{d}, $N=4$. In these panels, symbols '+' and '-' represent polarities of the phase singularity elements. Panels share identical scale bar: 10 $\mu$m. \textbf{e}-\textbf{h}, Evolution trajectories of these elements in three dimension, for the configurations corresponding to panels \textbf{a}-\textbf{d}. One of these trajectories for the element indexed as 1 is highlighted (here curve and data points represent theory and experiment, respectively). \textbf{i}-\textbf{l}, Projections of the trajectories in panels \textbf{e}-\textbf{h} onto the transverse plane.}
\end{figure} \newline
\indent The dependence of $\mathcal{E}_{N}$ on $N$ reveals that the increase factor shows an oscillating upward trend with the order $N$. When $N$ is an odd number, the relationship is reduced to $\mathcal{E}_{N}=N_\text{o}$; otherwise, it is expressed as $\mathcal{E}_{N}=N_\text{e}/2$ (here $N_\text{o}$ and $N_\text{e}$ denotes the odd and even integer numbers, respectively). Figure 6k showcases these relations: in both cases $\mathcal{E}_{N}$ grows linearly with $N$; however, $\mathcal{E}_{N}$ increases more rapidly for the odd-fold structures than that for their even-fold counterparts. To illustrate this unconventional oscillation, we first address the five-fold ($N=5$) rotationally symmetric structure, with radius $r_{0}=17\ \mathrm{\mu }$m and the angular offset $\theta _{0}=\pi /5$ (Fig. 6g). The theoretically propagated structure is displayed, at different values of $z$, as shown in Fig. 6g-6i, indicating that the rotating angle is $\psi_{5} = 6\pi/5$ after one-pitch propagation. The insets illustrate the theoretically recorded intensity patterns, with symbol '+' indicating the position of the first element (indexed as 1). With an even number $N$ (we address the typical case of $N=4$, see Fig. 6d-6f), the pivot alternates between positions $1$ and $3$, but it cannot reach positions at $2$ and $4$ after one-pitch propagation. The inserts in these panels depict the corresponding intensity patterns. In this case, it produces a rotating angle of $\psi_{4}=\pi$.  \newline
\indent The above predictions were implemented in the experiments. As mentioned, this requires to precisely derive the Fourier transform of the circular VAV chain, which includes an essential coupling term (see Methods). Using the setup (Fig. 3a), the phase singularities with desired VAV structures are generated, featured by $N=3, 4$ and $5$, see Fig. 7a-7c, respectively. In the case of $N=3$ (Fig. 7a), the experimentally observed period-increase factor is indeed three times the fundamental pitch ($\mathcal{E}_{3}=3$). Evolution trajectories up to a distance of $z=3p$ of all these elements are plotted in Fig. 7e, with data points and solid curves depicting the experimental and theoretical results, respectively. Detailed information on obtaining the theoretical trajectories refers to the Methods. Among them, the trajectory of the outer vortex labeled by $1$ is highlighted. The projection of the trajectories onto the transverse plane features a closed cogwheel shape, as shown in Fig. 7i, indicating a full period of evolution after $z$ approaches to $3p$. In the cases of $N=4$ and $5$, the predicted period-increase factors are $\mathcal{E}_{4}=2$ and $\mathcal{E}_{5}=5$, respectively. The experimental results confirm these predictions too. In particular, Fig. 7f demonstrates that the rotation period in the four-fold VAV chain is indeed twice the fundamental pitch, with the projected trajectories featuring a rounded square, see Fig. 7j, which is different from other cases with odd-fold structures. For the five-fold structure, the period becomes five times the fundamental pitch, see Fig. 7g, with the corresponding projected trajectories exhibiting a more intricate shape, see Fig. 7k. The experimental results show a full agreement with the theoretical predictions. We emphasize that the period-increase phenomenon is not limited to the studied VAV structure. Our theoretical model suggests that it is a general characteristic of rotationally symmetric VAV structures. For instance, a double-layer VAV structure, having a rotational symmetric distribution, exhibits similar tunable oscillations in the HOP. Figures S5 and S6 in section E of the supplementary present addtional observation of the period increase in the oscillation, manifested by the measured VAV light fields at different propagation distances. The measured period increase data was recorded, as indicated in Figs. 6j and 6k. \\
\indent As a reference, we further illustrate the evolution dynamics of five vortices with identical topological charge, symmetrically distributed in a circle of the same radius, while replacing the antivortex at the center with the vortex, see Fig. 7d. In this structure, owing to the lacking of the VAV couplings, the outer elements evolve independently in a conventional manner, showing harmonic oscillations with identical period of one pitch, as manifested by the three-dimensional trajectories of these elements (Fig. 7h) and their projections in the transverse plane (Fig. 7l). Figures S1c and S1d further confirm this trivial orbital evolution, see section A in the supplementary. In this case, the closed evolution trajectory forms a circle, featuring the usual harmonic oscillation. These findings in Fig. 7 (together with Figs. S5 and S6) highlight the nontrivial effects of the nonlocal VAV couplings on the vortex and antivortex motions in the HOP. \newline
\noindent \textbf{Discussion and outlooks}\newline
\noindent {In conclusion, we have revealed both theoretically and experimentally abnormal motion of the spatially structured wavepacket carrying a mix of vortex and antivortex wavefronts in the HOP. Unlike previously observed harmonic oscillations of the wavepackets \cite{Zhang2015,Moulieras2012,Jia2023}, the inclusion of the unique VAV couplings in the structured wavepacket causes a significant modification of their transverse motions, resulting in a nontrivial change to their original harmonic oscillations, which substantially modify the conventional motion trajectories. We look further insight into physical picture of the abnormal effects, by deriving a velocity model of the individual elements in the HOP system, see Sec. F in the supplementary. Based on this velocity model, we closely reveal how these VAV structures evolve in the HOP system, e.g., see Fig. S7 for the equilibrium VAV state and the four-fold symmetric VAV state.} \\
\indent Our investigation of the spatial VAV couplings under the action of the harmonic potential can be extended to the spatiotemporal domain. The spatiotemporal vortices of light, featuring transverse OAM, have received increasing attention recently \cite{Chong2020,Wan2023}. Particularly, a recent work has shown that, if transverse spatiotemporal vortex and antivortex are juxtaposed, they interact, as featured by their mutually perpendicular intersecting phase axes \cite{WanZhan2022}. As a result, the vortex and antivortex change their propagation directions in the spatiotemporal domain, which demonstrates the interaction between them. If the wavepackets, that carry the nontrivial transverse OAM, evolve in the HOP, abnormal motion of the wavepackets may be anticipated, due to the interaction between their constituent components. \\
\indent {The reported abnormal phenomena can find potential applications. As mentioned above, the VAV circular chains exhibit an explicit relation between order $N$ of the rotational symmetry and period increase $\mathcal{E}$. This unique one-to-one relation allows us to encode data according to the following mechanism. An applicable scheme has been illustrated in Fig. S8, see Sec. G of the supplmentary. This scheme is implementable in the experiment, providing a particular option for data encoding based on the well-structured braid and period adjustability in the HOP. We note that, the use of vortex knots and links as data carriers has been proposed recently \cite{Kong2022}. In comparison, our scheme can be applied to VAV light with defective structure. Such unique advantage can be attributed to the self-healing property of the VAV interaction in the HOP. In addition, our scheme can be applied in noisy environment, showing its robustness to the external noise.} \\
\indent The reported findings suggest other opportunities to explore unconventional dynamics of the VAV-coupled wavepackets with other different waveforms, which may be completely different from the conventional regimes. In particular, one can create complex topological VAV ensembles, such as vortex braids \cite{Dennis2010,Filippi2020,Zhong2021} in the HOP, with selected periodicities. Another intriguing option is to modify the potential, e.g., considering a two-dimensional anisotropic HOP, in which the particle's motion remains integrable, as the potential is separable. On the other hand, if the oscillation frequencies acting in the $x$ and $y$ directions are incommensurable, the corresponding trajectories are not Lissajous figures, but quasiperiodic ones densely covering the $\left( x,y\right) $ plane. \\
\indent Since vortices (and antivortices) are carried by waves in various physical settings, our investigations of the VAV interactions motivate further studies in those fields where waves are the main species of excitations, including electronics \cite{Verbeeck2010}, acoustics \cite{Zou2020}, hydrodynamics \cite{Wang2025}, and BEC \cite{Freilich2010,Croszek2016,Ferlaino2022}. Although the wave dynamics in different fields are governed by different physical laws, the propagation of vortex and antivortex modes is described by similar wave equations, resulting in shared phenomenology \cite{Dudley2014,Rozenman2019,Malomed2022}. For example, various kinds of topological waveforms including the vortices and antivortices have been generated on the water surface \cite{Wang2025}. The topological dynamics is described by the water-wave equation, which is similar to the Schr$\ddot{\text{o}}$dinger equation in optics \cite{Dudley2014,Rozenman2019}.  \\
\indent We finally emphasize that while numerous techniques have been employed for structured light manipulations \cite{Forbes2021,Shen2019}, the optical VAV couplings remain largely overlooked, particularly in the experiment. Our demonstration of the intrinsic coupling effects in the HOP provide a distinct approach to manipulate the structured wavepackets. We believe that the VAV couplings may become an important mechanism for the design and control of many kinds of wavepacket, both in the spatial and time domains.  
\newpage
\noindent\textbf{Methods}\newline
\noindent \textbf{Theoretical model for the vortex (antivortex) motion without VAV couplings}\newline
\noindent We study a vortex (antivortex) wavepacket in the HOP, in the absence of the VAV couplings, namely, it only comprises the vortex (or antivortex) pivots with identical vorticity. Without loss of generality, we assume that the wavepacket contains $N$ vortices at the propagation distance of $z=0$. Mathematically, it can be expressed as a form of 
\begin{equation}\label{G}
	G\left( {x,y}, z=0 \right) = E\left( {x,y} \right) \times p\left( u \right)
\end{equation}
where $p(u) = \sum\nolimits_{n = 0}^N {{a_n}{u^n}}$ is a complex polynomial function of complex variable $u=x+iy$, with its coefficients $a_n$ determining the spatial arrangement of the vortex pivots in the Gaussian envelope $E(x,y)$. Equation (8) shows that the Gaussian envelope is polynomially modulated in two dimensional space, producing multiple phase singularities at positions where the roots of the function $p(u)$ locate. $p(u)$ can be rewritten as another form of: $p(u) = \prod\nolimits_{n = 1}^N {\left( {u - {c_n}} \right)} $, where $c_n=x_+^{(n)}(0)+iy_+^{(n)}(0)$ denotes its $n^{\text{th}}$ root, standing for the transverse location of the $n^{\text{th}}$ pivot at a propagation distance of $z=0$. \\
\indent We then investigate propagation dynamics of this vortex configuration in the HOP, described by the linear Sch\"{o}dinger equation [Eq. (3)]. Generally, the solution of Eq. (3) initiated by Eq. (8) can be derived out by means of the Green function \cite{Simon1994,Zhang2015}. In terms of the scaled Cartesian coordinate system, $(\xi,\eta  = (\sqrt {{n_0}{k_0}\alpha } x, \sqrt {{n_0}{k_0}\alpha } y)$, the solution can be expressed as
\begin{equation}
\begin{gathered}
G\left( {x , y ,z} \right) = \frac{ - i\csc \left( {\alpha z} \right)\exp\left[i\cot(\alpha z)(\xi^2 +\eta^2)\right]}{{2{\pi ^2}}} \hfill \\
\times\iint {{G\left( {\frac{{\xi '}}{{\sqrt {{n_0}{k_0}\alpha } }},\frac{{\eta '}}{{\sqrt {{n_0}{k_0}\alpha } }}},0 \right)}\exp\left[{i\cot ({\alpha z})( {{{\xi '}^2} + {{\eta '}^2}}) - i\csc ( {\alpha z})( {\xi \xi ' + \eta \eta '})}\right]d\xi 'd\eta '} \hfill \\
\end{gathered}
\end{equation}
Since the initial waveform $G(x,y,z=0)$ is polynomially modulated, equation (9) can be represented as a novel waveform, which is modulated by a propagated polynomial function [different from $p(u)$]. In this scenario, the propagation dynamics of the configuration is analytically expressed as the following form:
\begin{equation}\label{Solution}
G\left( {x,y,z} \right) = E\left( {x,y,z} \right) \times \prod\limits_{n = 1}^N {\left( {Tu - {c_n}} \right)} 
\end{equation}
where $T = {\left( {\cos \tau  + i\sigma \sin \tau } \right)^{ - 1}}$, as defined in the main text. It becomes a fundamentally essential parameter that uniquely governs the pivots' motions in the HOP. \\
\indent We carefully extend the theoretical framework to the anti-configuration, which comprises multiple antivortices. In this anti-configuration, a solution can be also obtained from the Sch\"{o}dinger equation [Eq. (3)], written as:
\begin{equation}
{G}\left( {x,y,z} \right) = E\left( {x,y,z} \right) \times \prod\limits_{n = 1}^N {\left( {T{u^*} - d_n^*} \right)},
\end{equation}
where $d_n=x_-^{(n)}(0)+iy_-^{(n)}(0)$ denotes the initial location of the $n^{\text{th}}$ antivortex pivot. Evidently, in both cases [Eqs. (10) and (11)], the initial positions of the vortices and the antivortices are considerably influenced by the propagation-varying parameter $T$. \\
\indent To monitor their locations in the course of propagation, we equate the polynomially-modulated light field $G(x,y,z)$ to zero (for both configurations), namely, the value of the propagated polynomial functions equates to zero: $Tu - {c_n} = 0$ for the vortex configuration; and $Tu^* - {d_m^*} = 0$ for the antivortex configuration. According to these conditions, dynamical locations of the individual vortex and antivortex elements can be analytically derived as
\begin{equation}
	\begin{gathered}
		x_+^{(n)}(z) = {x_+^{(n)}}(0)\cdot\cos \tau  - {y_+^{(n)}}(0)\sigma\cdot\sin \tau \hfill \\
		y_+^{(n)}(z) = {x_+^{(n)}}(0)\sigma\cdot \sin \tau  + {y_+^{(n)}}(0)\cdot\cos \tau \hfill \\
	\end{gathered}
\end{equation}
and
\begin{equation}
	\begin{gathered}
		x_-^{(n)}(z) = {x_-^{(n)}}(0)\cdot\cos \tau  + {y_-^{(n)}}(0)\sigma \cdot \sin \tau \hfill\\
		y_-^{(n)}(z) =  - {x_-^{(n)}}(0)\sigma \cdot \sin \tau  + {y_-^{(n)}}(0)\cdot\cos \tau \hfill \\
	\end{gathered}
\end{equation}
respectively. Equations (12) and (13) suggest that, in the absence of the VAV couplings, the elements exhibit independent propagation in the HOP, analogous to the conventional harmonic oscillations in the two dimensional space. This motion dynamics is captured by a rotational transformation matrix, derived from Eqs. (12) and (13). To this end, we rewrite Eqs. (12) and (13) in a matrix form as: 
\begin{equation}
{\mathbf{r}_{\pm}^{n} }\left( z \right) = {\mathbf{V}_{\pm}}(\tau){\mathbf{r}_{\pm}^{n}}\left( 0 \right)
\end{equation}
where $\mathbf{r}_{\pm}^{n}$ denotes the position vectors of the $n^{\text{th}}$ vortices ($+$) and antivortices ($-$). The rotational transformation matrix $V_{\pm})(\tau)$ is therefore written as
\begin{equation}
{\mathbf{V}_ \pm }(\tau) = \left[ {\cos \tau, \mp \sigma \sin \tau; \pm \sigma \sin \tau,\cos \tau} \right]
\end{equation}
Here $\mathbf{V}_+(\tau)$ and $\mathbf{V}_-(\tau)$ represents the transformation matrix for the vortex and antivortex elements in the configurations, respectively. \\
\noindent \textbf{Derivation for the coupling terms in the HOP} \\
\noindent  We provide a detailed derivation for the VAV and light-potential couplings. We start from the term 
\begin{equation}
B=\frac{2i\sin(\tau)(n_0k_0\alpha)^{-1} }{{\cos \tau +i\sigma \sin \tau }}
\end{equation}
We assume that this term can be rewritten as a sum of two, $B=B_0+B_1 $, where $B_0$ and $B_1$ denote, respectively, the intrinsic VAV coupling and the light-potential coupling. To derive these coupling terms, we consider the Taylor expansions of $\sin\tau$ and $\cos\tau$, written as $\sin\tau=\tau-\tau^3/6+...$ and $\cos\tau=1-\tau^2/2+...$, respectively. We substitute these expansions in $B$, yielding 
\begin{equation}
B=\frac{2iw_0^2z+2iw_0^2(-\alpha^2z^3/6+...)}{(n_0k_0w_0^2+i2z)+n_0k_0w_0^2[(-\tau^2/2+...)+i\sigma(-\tau^3/6+...)]}
\end{equation}
Using the identity: $1/(n+m)=1/n-m/[n(n+m)]$, equation (17) can be re-expressed as follows, after some algebraic manipulations:
\begin{equation}
B=\frac{2iw_0^2z}{n_0k_0w_0^2+2iz}+\frac{2i[w_0^2+2iz/(n_0k_0)](-\alpha^2z^3/6+...)+iw_0^2z(-\sigma\alpha z^2-\alpha^2z^3/6+...)}{(n_0k_0w_0^2+2iz)(\cos\tau+i\sigma\sin\tau)}
\end{equation}
{The first term of Eq. (18) depends on $z$ only, dominating the VAV coupling:
\begin{equation}
B_0=\frac{2iw_0^2z}{n_0k_0w_0^2+2iz}
\end{equation}
The second term of Eq. (18) consists of an infinite number of terms with respect to potential depth $\alpha$, which represents light-potential coupling, and thus it is denoted as $B_1$. We find that this infinite summation is convergent, whose explicit expression takes the form of $B_1=B-B_0$, which yields
\begin{equation}
B_1=\frac{2iw_0^2(\alpha^{-1}\sin\tau-z\cos\tau)}{(n_0k_0w_0^2+2iz)(\cos\tau+i\sigma\sin\tau)}
\end{equation}}\\
\noindent\textbf{Design of the phase-only hologram} \\
{The challenges for observing the abnormal VAV phenomena are two-fold. From the theoretical level, one has to develop a theoretical model that closely governs the VAV coupling in the system. This is a rather complicated many-body interaction process, which involves cumbersome algebra to figure out the many-body coupling term. From the experimental level, the experimental technique closely depends on such a theoretical coupling model in the design of the VAV structured light. We address these issues, and experimentally generate the VAV-coupled wavepackets, allowing to reveal the underlying abnormal phenomena in the HOP. This requires to accurately encode both the phase and amplitude information of the desired VAV configuration. We mention that, previous encoding techniques addressed solely the uncoupled term, while neglecting the coupling term of the VAV wavepacket \cite{Ferrando2023,Fan2021,Gao2021,Ding2018,Vyas2007}, as specified below. In this section, we theoretically derive these essential coupling terms, which allow us, for the first time, to develop the holographic method for observing the VAV interactions in the HOP.} \\
\indent We study two characteristic VAV-coupled wavepackets in the HOP. For the case of the fundamental VAV dipole, the Fourier transform is expressed as
\begin{equation}\label{eq:FTG}
	\tilde G\left( {{v_x},{v_y}} \right) = {{\tilde E}_0}\left( {{v_x},{v_y}} \right) \times \left[ {\left( {\tilde Tv - d/2} \right)\left( {\tilde T{v^*} + d/2} \right) + w_0^2} \right]
\end{equation}
where $\tilde E_0\left( {{v_x},{v_y}} \right)\!\! =\!\! {w_0^2}/2\exp \left({ - {w_0^2}\left( {v_x^2 + v_y^2} \right)/4} \right)$ represents the Fourier transform of the Gaussian host field, as defined in the main text. In this expression, $v= v_x+iv_y$ denotes a complex variable in the frequency domain, corresponding to the spatial-domain variable $u=x+iy$. Here $\tilde T = - iw_0^2/2$. Equation (21) comprises two essential terms, the $\tilde{T}$-related one $\tilde{F}_0=\left( {\tilde Tv - d/2} \right)\left( {\tilde T{v^*} + d/2} \right)$ and $\tilde{F}_c=w_0^2$, demonstrating a form similar to the VAV dipole solution in Eq. (4). We emphasize that the Fourier-domain coupling term featured by $\tilde{F}_c$ should be included in the design of the phase-only hologram. \\
\indent For the more involved VAV circular chain, which comprises elements symmetrically distributed in a circle of radius $r_0$ around the center (as illustrated in Fig. 6), we successfully derive its Fourier transform, expressed as 
\begin{equation}
	\tilde G\left( {{v_x},{v_y}} \right) = {{\tilde E}_0}\left( {{v_x},{v_y}} \right) \times \left\{ {\left( {\tilde T{v^*}} \right)\left[ {{{\left( {\tilde Tv} \right)}^N} - r_0^N{e^{iN{\theta _0}}}} \right] + w_0^2N{{\left( {\tilde Tv} \right)}^{N - 1}}} \right\}
\end{equation}
Equation (22) exhibits a mathematical form similar to the expression in Eq. (7). It also includes two important terms: one featuring the geometrical configuration of this wavepacket, as suggested by $\tilde{F}_0=\left( {\tilde T{v^*}} \right)\left[ {{{\left( {\tilde Tv} \right)}^N} - r_0^N{e^{iN{\theta _0}}}} \right]$ and the other term featuring the couplings among these vortex and antivortex elements, as suggested by $\tilde{F}_c=w_0^2N{{\left( {\tilde Tv} \right)}^{N - 1}}$. \\
\indent {To generate the expected VAV wavepackets, we employ a reliable phase-only encoding technique as presented in Ref. \cite{Bolduc2013}. Based on Eqs. (21) and (22), the Fourier transforms of the VAV-coupled wavepackets can be equivalently expressed in the following form, based on the Euler's formula
\begin{equation}
\tilde G\left( {{v_x},{v_y}} \right) =\tilde{E}_0(\tilde{F}_0+\tilde{F}_c)= {{\tilde G}_A}\left( {{v_x},{v_y}} \right) \times \exp \left[ {i\tilde \theta \left( {{v_x},{v_y}} \right)} \right]
\end{equation}
where ${{\tilde G}_A}\left( {{v_x},{v_y}} \right)$ and $\tilde \theta \left( {{v_x},{v_y}} \right)$ represent the amplitude and phase Fourier distributions of $\tilde G\left( {{v_x},{v_y}} \right)$, respectively.  The phase-only hologram loaded onto the spatial light modulator (SLM, see the experimental setup) can be therefore expressed as \cite{Bolduc2013}
\begin{equation}
\tilde H\left( {{v_x},{v_y}} \right) = \tilde M\left( {{v_x},{v_y}} \right) \times \text{Mod}\left[ {\tilde \Phi \left( {{v_x},{v_y}} \right) + \frac{{2\pi {v_x}}}{\Lambda },2\pi } \right]
\end{equation}
where $\tilde M\left( {{v_x},{v_y}} \right)$ and $\tilde \Phi \left( {{v_x},{v_y}} \right)$, which include information of the coupling term, are closely relevant to the amplitude and phase information of the target VAV wavepacket. Specifically, they are expressed as: $\tilde M\left( {{v_x},{v_y}} \right) = 1 + {\text{sin}}{{\text{c}}^{ - 1}}[ {{{\tilde G}_A}\left( {{v_x},{v_y}} \right)} ]/\pi $ and $\tilde \Phi \left( {{v_x},{v_y}} \right) = \tilde \theta \left( {{v_x},{v_y}} \right) - \pi \tilde M\left( {{v_x},{v_y}} \right)$, respectively. The sinc function in $\tilde{M}$ is defined as $\text{sinc}(\cdot)= \sin(x)/x$ for $x \in [0, \pi]$. The parameter $\Lambda$ in Eq. (24) denotes a blazed grating period, which is used to diffract the desired light field into the first-order diffraction component of the hologram at the focal plane of Fourier lens O1. In our experiments, the period of the one-dimensional blazed grating is set to 64 $\mu$m.}\\
\indent Additional experimental details are provided in the following. First, the VAV coupling term $\tilde{F}_c$, which is nested in the complete light field $G(x,y)$, must be taken into account in the calculations of $\tilde M\left( {{v_x},{v_y}} \right)$ and ${\tilde \Phi \left( {{v_x},{v_y}} \right)}$. Second, the presented encoding technique requires that the light beam incident onto the SLM should be a plane wave. To realize this condition, the incident light beam should be appropriately expanded using the telescope system (L1 and L2) to cover the entire SLM screen, effectively approximating a plane wave. Otherwise, the obtained phase-only hologram for the VAV wavepackets is less accurate, which degrades the quality of the VAV coupling. Third, to numerically implement the inversion of the sinc function, the amplitude ${\tilde G}_A$ should be normalized to the domain $[0,1]$, while maintaining its Fourier distribution unchanged. Following these essential details, the expected VAV-coupled wavepackets can be reliably encoded into the phase-only holograms and generated experimentally, facilitating our studies of the nontrivial VAV interactions in the HOP. \\
\noindent \textbf{Evolution trajectories of the pivots in rotationally symmetric VAV-coupled structure} \newline
\noindent To further analyze the motion of pivots of the vortices and antivortices in the rotationally symmetric VAV wavepacket, we theoretically extract their coordinates at different propagation distances, by equating the real and imaginary parts of the complete light field described in Eq. (7) to zero. Without loss of generality, we consider the evolution trajectory of the $n^{\text{th}}$ vortex pivot in the structure and represent its transverse location in the polar coordinate system. This leads to the following polynomial equation, where the propagation-varying radial coordinate $r(z)$ of the $n^{\text{th}}$ pivot satisfies:
\begin{equation}
{\gamma _{1}} r^{2N}+{\gamma _{2}}r^{2\left( {N-1}\right) }+{\gamma _{3}}r^{2\left( {N-2}\right) }=r_{0}^{2N} \label{eq:radial}
\end{equation}%
where ${\gamma _{1}}={\left\vert T\right\vert ^{2N}}$, ${\gamma _{2}}=2{\left\vert T\right\vert ^{2\left( {N-1}\right) }}\mathrm{Re}[{NBT{{\left( {{T^{\ast }}}\right) }^{-1}}}]$ [here Re($\cdot$) is used to extract the real part and $*$ denotes the complex operator], and ${\gamma _{3}}={\left( {N\left\vert B\right\vert {{\left\vert T\right\vert}^{N-2}}}\right) ^{2}}$ (recall $r_{0}$ is the radius of the circular VAV chain). These three coefficients $\gamma_1$, $\gamma_2$ and $\gamma_3$ are dependent on the propagation distance. On the other hand, the propagation-varying azimuthal coordinate of the $n^{\text{th}}$ pivot is written as
\begin{equation}
{\phi _n}(z) =\theta _0 - \frac{{\arg \left\{ {{{\left( {Tr} \right)}^N} + NB{{\left( {Tr} \right)}^{N - 2}}} \right\} - 2\left( {n - 1} \right)\pi }}{N}
\end{equation}
where $\arg \{\}$ denotes the angle of the complex expression, and $n=1$, 2, $ \cdots$, $N$. Here $\theta_0$ is an initial angle offset of the VAV configuration. These two equations (25, 26) suggest that, under the action of the VAV couplings, the pivots' motion no longer follows the Lissajous trajectories, exhibiting new remarkable patterns. By solving Eqs. (25) and (26), we can obtain the exact evolution trajectory of the individual elements in the VAV configuration. \\
\indent In the following, we demonstrate an example of solving the evolution trajectories [based on the Eqs. (25) and (26)], in the case of the three-fold ($N=3$) rotationally symmetric VAV structure. In this scenario, the radial coordinate, according to the Eq. (25), can be expressed as a six-order polynomial equation, written as follow,
\begin{equation}
{\gamma _1}r{^6} + {\gamma _2}r{^4} + {\gamma _3}r{^2} = r_0^6
\end{equation}
where ${\gamma _1} = |T|^6$, ${\gamma _2} = 6{\mathop{\rm Re}\nolimits}[ {B{T^*}{T^{ - 1}}}]|T|^4$, and ${\gamma _3} = 9{\left| B \right|^2}|T|^2$. The six-order polynomial equation (27) can be rewritten as the relatively simpler cubic one, by taking advantage of the relationship: $X=|T|^2 r^2 + |T|^{-4}\gamma_2 / 3$. The simplified equation becomes
\begin{equation}
{X^3} + {\Gamma _1}X + {\Gamma _2} = 0
\end{equation}
where 
\begin{equation}
	\begin{gathered}
		{\Gamma _1} = 9{\gamma _3}{\left| T \right|^{ - 2}} - 12{\gamma _2}{\left| T \right|^{ - 4}} \hfill \\
		{\Gamma _2} = 8{\left( {{\gamma _2}{{\left| T \right|}^{ - 4}}} \right)^3} - 18{\gamma _3}{\gamma _2}{\left| T \right|^{ - 6}} - r_0^6 \hfill \\
	\end{gathered}
\end{equation}
Implementing the Cardano's formula, the root of the cubic equation (28) can be expressed as 
\begin{equation}
	X = {\left( { - \frac{{{\Gamma _2}}}{2} + {\chi ^{\frac{1}{2}}}} \right)^{\frac{1}{3}}} + {\left( { - \frac{{{\Gamma _2}}}{2} - {\chi ^{\frac{1}{2}}}} \right)^{\frac{1}{3}}}
\end{equation}
where $\chi  = {\left( {\frac{{{\Gamma _1}}}{3}} \right)^3} + {\left( {\frac{{{\Gamma _2}}}{2}} \right)^2}$. As a result, the radial coordinates of the outer pivot has the form of 
\begin{equation}
	{r\left( z\right)} = {\left| T \right|^{ - 1}}{\left( {X - {{\left| T \right|}^{ - 4}}{\gamma _2}/3} \right)^{\frac{1}{2}}}
\end{equation}
It is evident that the outer elements follow identical radial coordinate, described by Eq. (31). On the other hand, the azimuthal coordinate for the $n^{\text{th}}$ pivot in the three-fold VAV configuration is achieved as
\begin{equation}
{\phi _n}\left( z \right) = {\theta _0} - \frac{{\arg \left\{ {{{\left( {T{r}} \right)}^3} + 3B\left( {T{r}} \right)} \right\} - 2\left( {n - 1} \right)\pi }}{3}
\end{equation}
with the integer number $n=1$, 2 and 3, corresponding to the outer vortex pivots. The combination of Eqs. (31) and (32) reveals that the outer elements exhibits similar evolution trajectories, differing by their initial offset $\theta_0$. Thus, the projection of these trajectories on the transverse plane exhibits rotational symmetry (as evident in Fig. 7). Using this framework, one can obtain trajectories of the individual elements in the higher-order vortex-antivortex structures, as illustrated in Fig. 7 of the main text, for the cases of $N=3$, $N=4$, and $N=5$. \\
\noindent\textbf{Data availability}\\
\noindent All data that supports the plots presented in this paper and other findings of this study are available from the corresponding authors upon request.\\
\noindent \textbf{Code availability} \\
\noindent The custom code used in this study is available from the corresponding authors upon request. \\
 
\noindent \textbf{Acknowledgements}\newline
\noindent This study was supported by the National Key Research and Development Program of China (grant no. 2022YFA1404800 to Y.C.), the National Natural Science Foundation of China (grant nos. 12522414 and 12374306 to S.F., 12192254 and 12534014 to Y.C., 12504382 to H. L.), and the Guangdong Basic and Applied Basic Research Foundation (grant no. 2025A1515011694 to H. L.). \newline
\noindent \textbf{Author Contributions}\newline
\noindent S. Fu conceived the concept. S. Fu, H. Lin, J. Jia, B. Malomed and Y. Cai carried out the analytical considerations. S. Fu, and H. Lin drafted the paper. S. Fu, B. Malomed and Y. Cai revised the paper. S. Fu, H. Lin, and J. Jia performed the experiments, S. Fu, H. Lin, C. Liang, Y. Hu, and Y. Cai participated in the design of the phase-only hologram and performed the analysis. H. Lin, J. Jia and C. Liang performed numerical simulations. S. Fu and Y. Cai supervised the project. All authors participated in discussions and contributed to the editing of the article. \newline
\noindent \textbf{Conflicting interests}\newline
\noindent The authors declare no competing interests.
\end{document}